\documentclass[9pt,conference]{IEEEtran}
\IEEEoverridecommandlockouts

\usepackage{cite}
\usepackage{amsmath,amssymb,amsfonts}
\usepackage{algorithmic}
\usepackage{graphicx}
\usepackage{textcomp}
\usepackage{xcolor}
\usepackage{multirow}
\usepackage{booktabs}
\usepackage{marvosym}
\usepackage{cite}
\usepackage{setspace} 
\usepackage{ulem}

\def\BibTeX{{\rm B\kern-.05em{\sc i\kern-.025em b}\kern-.08em
    T\kern-.1667em\lower.7ex\hbox{E}\kern-.125emX}}
\begin{document}



\def\x{{\mathbf x}}
\def\L{{\cal L}}

\title{ZipEnhancer: Dual-Path Down-Up Sampling-based Zipformer for Monaural Speech Enhancement \\
}

\author{

\IEEEauthorblockN{1\textsuperscript{st} Haoxu Wang}
\IEEEauthorblockA{\textit{Speech Lab} \\
\textit{Alibaba Group}\\
China \\
wanghaoxu.whx@alibaba-inc.com}

\and

\IEEEauthorblockN{2\textsuperscript{nd} Biao Tian}
\IEEEauthorblockA{\textit{Speech Lab} \\
\textit{Alibaba Group}\\
China \\
tianbiao.tb@alibaba-inc.com}


\and
}

\maketitle

\begin{abstract}
In contrast to other sequence tasks modeling hidden layer features with three axes, Dual-Path time and time-frequency domain speech enhancement models are effective and have low parameters but are computationally demanding due to their hidden layer features with four axes. We propose ZipEnhancer, which is Dual-Path Down-Up Sampling-based Zipformer for Monaural Speech Enhancement, incorporating time and frequency domain Down-Up sampling to reduce computational costs. We introduce the ZipformerBlock as the core block and propose the design of the Dual-Path DownSampleStacks that symmetrically scale down and scale up. Also, we introduce the ScaleAdam optimizer and Eden learning rate scheduler to improve the performance further. Our model achieves new state-of-the-art results on the DNS 2020 Challenge and Voicebank+DEMAND datasets, with a perceptual evaluation of speech quality (PESQ) of 3.69 and 3.63, using 2.04M parameters and 62.41G FLOPS, outperforming other methods with similar complexity levels.

\end{abstract}


\section{Introduction}

Speech enhancement (SE) aims to improve natural speech signals captured with devices often degraded by complex acoustic scenarios, filtering out noise and boosting clarity for better understanding.
By enhancing speech signals, this technology contributes to the efficiency of communication in real-world applications.
In current SE methods, there are generally two categories: time domain SE and time frequency domain (TF) SE, and use various networks, including recurring neural network (RNN)\cite{CRNN,ResidualRNN}, Transformer\cite{AttentionBaseSE,SEConformer,TimeSESelfatt}. 
Time-domain models often encode the noisy waveform into hidden-layer features, which are then processed by a Transformer-based encoder, and then restore the estimated waveform\cite{SEGAN,TCNN,RealtimeSE,SEConformer,TimeSESelfatt}. 
TF domain methods use neural networks to predict clean short-time Fourier transform (STFT) features from noisy audio-generated STFT features and reconstruct the waveforms using inverse STFT (ISTFT)\cite{FRCRN,TF1,TF2,DPTFSNet,TridentSE}. To enhance audio quality in low signal-to-noise ratio (SNR) scenarios, some TF domain methods implicitly or explicitly predict clean STFT phase information using complex, magnitude, and phase information\cite{DBAIAT,CMGAN,mpsenet} and achieve good results in SE tasks. 


\begin{figure*}[!t]
	\centering
	\includegraphics[scale=0.75]{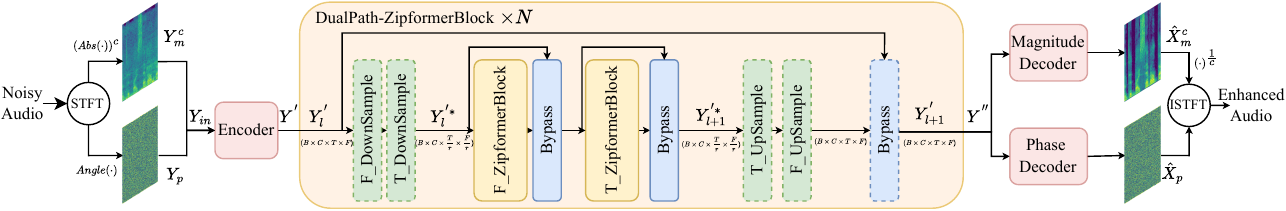}
        \vspace{-10pt}
	\caption{
		The overall architecture of our ZipEnhancer.
	}
	\label{fig:model}
        \vspace{-10pt}
\end{figure*}

\begin{figure*}[!t]
	\centering
	\includegraphics[scale=0.75]{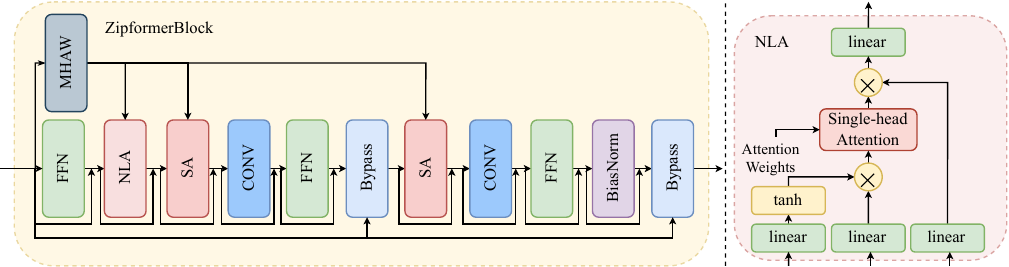}
        \vspace{-10pt}
	\caption{
		(Left): The structure of T/F-ZipformerBlock. (Right): The structure of Non-Linear Attention module.
	}
	\label{fig:zipformerblock}
        \vspace{-18pt}
\end{figure*}

Unlike conventional sequence tasks such as Automatic Speech Recognition (ASR)\cite{attentionallyouneed,transfomer1,transformer2}, which model hidden features $X \in R^{B \times T \times C}$ on the time dimension, researchers in speech separation (SS) \cite{dprnn,sepformer,tfgridnet,tflocoformer} and SE\cite{CMGAN,TridentSE,mpsenet,USES,tflocoformer,CADBConformer} retain an additional frequency dimension, constructing hidden features $(B, T, F, C)$ with four axes. They commonly use RNN or Transformer-style Dual-Path modeling on the time $T$ and frequency $F$ dimensions. Despite SS and SE models having a parameter count of several million (M), much smaller than ASR, their computational costs remain high due to Dual-Path modeling. The model's parameter quantity does not fully reflect its modeling capability\cite{ComplexityScaling,zhang24i_interspeech}.
Recent ASR research has made efforts to develop efficient Encoders such as Efficient Conformer\cite{EfficientConformer}, Squeezeformer\cite{kim2022squeezeformer}, and Zipformer\cite{zipformer}. Efficient Conformer uses progressive downsampling, while Squeezeformer uses a U-Net structure to downsample the time resolution of intermediate hidden features to 12.5Hz. In comparison, Zipformer introduces varying downsampling ratios at different layer levels, proposing a more aggressive downsampled U-Net structure. This time domain downsampling effectively reduces the computational cost and enhances the ASR Encoder's efficiency. 
Inspired by these methods, we introduce time domain downsampling into the SE model, extending it to frequency domain downsampling to reduce computational costs.
Therefore, based on MP-SENet\cite{mpsenet}, we build ZipEnhancer, a TF-Domain SE model that incorporates different ratio downsampling of time-domain and frequency-domain features in Dual-Path Transformer-like blocks. The model includes an Encoder, Dual-Path ZipformerBlocks, and Decoders. The Encoder initially models magnitude and phase to obtain hidden-layer features. Subsequently, the Dual-Path ZipformerBlocks use DownSampleStacks to sequentially model the time and frequency domains, followed by the Magnitude Decoder restoring the magnitude spectrum and the Phase Decoder explicitly restoring the phase spectrum.
Compared to the TF-Conformers or TF-GRUTransformers in MP-SENet, we modify the dual-path Blocks to FT-ZipformerBlocks using Zipformer. Additionally, we propose DownSampleStacks with paired Downsample and Upsample structures, conducting symmetric scaling down and up in both time and frequency domains to reduce the computational cost and model temporal and frequency domain information at different resolution levels. Also, we introduce the ScaleAdam optimizer and the Eden scheduler to improve performance further.
Extensive experiments on the DNS Challenge 2020 (DNS2020) \cite{dns2020} and VoiceBank+DEMAND\cite{voicebankdemands} datasets demonstrate the superiority of our ZipEnhancer.
Our ZipEnhancer outperformed models of similar scale, achieving a new state-of-the-art (SOTA) perceptual evaluation of speech quality (PESQ) score of 3.69 on the DNS2020 dataset and 3.63 on the Voicebank+DEMAND dataset with 2.04M parameters and 62.41G floating-point operations per second (FLOPS).

\vspace{-3pt}
\section{The proposed ZipEnhancer}
\vspace{-3pt}

\subsection{The overall architecture of the model}
\vspace{-3pt}

Figure \ref{fig:model} shows the overall architecture of our model. We use ZipEnhancer with codec structure to estimate the clean signal $x \in R^{L}$ from the noisy audio signal $y = x + z \in R^{L}$, with $L$ representing the audio length. Initially, we extract the TF magnitude spectrum $Y_{m} \in R^{T \times F}$ and the wrapped phase spectrum $Y_{p} \in R^{T \times F}$ from the noisy signal y. We then perform power-law compression on the magnitude spectrum $Y_{m}$ using a scale factor c to obtain $Y_{m}^{c}$, and concatenate it with $Y_{p}$ along a new channel dimension to obtain the input feature $Y_{in}$ for the Encoder, in $R^{2 \times T \times F}$.

The Encoder uses convolutional modeling on $Y_{in}$ to obtain hidden-layer time-frequency features $Y^{'} \in R^{C \times T \times F^{'}}$. The stacked Dual-Path ZipformerBlocks alternately conduct sequential modeling on the time and frequency domains, capturing global and local acoustic information. Downsampling blocks aid in aggregating and restoring original length information while reducing the model's computational cost. Subsequently, we obtain new hidden-layer features $Y^{''} \in R^{C \times T \times F^{'}}$ for future decoding.
Parallel magnitude spectrum decoder and phase spectrum decoder reconstruct the estimated clean magnitude spectrum $\hat X_{m}$ and clean phase spectrum $\hat X_{p}$. The estimated enhanced waveform $\hat x$ is then reconstructed using ISTFT.

\vspace{-3pt}
\subsection{DualPathZipformerBlocks}
\vspace{-3pt}

\subsubsection{DownSampleStacks}

Unlike the Single-Path downsampling blocks in ASR, the Dual-Path structure for alternate modeling in the time-frequency domain can be designed to include modules that simultaneously downsample in both the time and frequency dimensions. As shown in Figure \ref{fig:model}, within the DownSampleStack, we use paired DownSample and UpSample modules to achieve symmetric scaling down and up in the time or frequency length. The module for Down-Upsampling in the time dimension is named T\_DownSample and T\_UpSample, while the module for Down-Upsampling in the frequency dimension is termed F\_DownSample and F\_UpSample. The sampling ratio is denoted as $r$. Our sampling approach follows the simplest method, as in Zipformer. For instance, with a factor of $r$, the DownSample module averages every $r$ frames using $r$ learnable scalar weights (after softmax normalization), and the UpSample module repeats each frame $r$ times. Through this downsampling process, the original feature $Y^{'}_{l} \in R^{C \times T \times F}$ at layer $l$ compresses to $Y^{'*}_{l} \in R^{C \times \frac{T}{r} \times \frac{F}{r}}$, which then feeds into the Dual-Path Zipformerblock for modeling the time and frequency domains at lower frame rates. Additionally, a Bypass module, similar to a residual connection (described in Sec. \ref{sec:bypass}), shortens the unsampled information to enhance the modeling capability of the model. Note that the downsampling structure of the dashed line in Figure \ref{fig:model} is optional and is deleted at $r = 1$.

\subsubsection{ZipformerBlock}
\label{sec:bypass}

In the Dual-Path structure, the F\_ZipformerBlock and T\_ZipformerBlock share the same model architecture. For the compressed $Y^{'*}_{l}$ within a mini-batch, with a shape of $B \times \frac{T}{r} \times \frac{F^{'}}{r} \times C$, where $B$ is the batch size, we organize it into $\frac{BT}{r} \times \frac{F^{'}}{r} \times C$, and input it into the F\_ZipformerBlock to model frequency domain correlations. Then, we organize it into $\frac{BF^{'}}{r} \times \frac{T}{r} \times C$ and input it into the T\_ZipformerBlock to model temporal correlations, ultimately obtaining the output $Y^{'*}_{l+1}$.

The standard ConformerBlock includes a pair of Feed-Forward Networks (FFN), Multi-Head Self-Attention (MHSA), Convolution (CONV), and LayerNorm (LN). ZipformerBlock includes various improvements compared to the ConformerBlock.
As shown in Figure \ref{fig:zipformerblock}, a ZipformerBlock comprises two Conformer-like blocks, differing by excluding LN and including only BiasNorm at the end. Also, the third FFN before the second MHSA is removed, leaving three FFNs. To further reduce computational costs, it uses a separate Multi-Head Attention Weight (MHAW) to compute attention weights for sequence global modeling, reusing these weights in Non-linear Attention (NLA) and two Self-Attention (SA) modules.
In the ZipformerBlock, the SA no longer recalculates attention weights but uses the weights provided by MHAW. Additionally, the ZipformerBlock introduces several Bypass modules for shortcut connections within the block.

\begin{table}[t]\centering
\small

\setlength{\tabcolsep}{1.4mm}{
  \centering
  \footnotesize
  \caption{\label{tab:config} {Different hyper-parameter configurations of the ZipEnhancer.
}}
  \begin{tabular}{ccccccccccccc}
  \toprule

  Model & N & Ratios & C & Heads & Para[M] & FLOPS[G] \\

  \midrule

  S & 4 & \{1, 2, 2, 1\} & 64 & 4 & 2.04 & 62.85 \\
  S2 & 4 & \{1, 1, 1, 1\} & 64 & 4 & 2.04 & 80.61 \\
  S3 & 4 & \{1, 2, 4, 1\} & 64 & 4 & 2.04 & 60.79 \\
  S4 & 4 & \{1, 2, 4, 2\} & 64 & 4 & 2.04 & 51.91 \\
  S5 & 4 & \{1, 4, 4, 2\} & 64 & 4 & 2.04 & 49.84 \\
  S6 & 4 & \{2, 3, 4, 2\} & 64 & 4 & 2.04 & 41.47 \\
  S7 & 4 & \{2, 6, 8, 2\} & 64 & 4 & 2.04 & 40.06 \\
  S8 & 4 & \{3, 6, 8, 3\} & 64 & 4 & 2.04 & 36.94 \\

  \midrule

  M & 6 & \{1, 2, 3, 4, 2, 1\} & 128 & 8 & 11.34 & 266.96 \\
  




  \bottomrule

\end{tabular}
}
\vspace{-18pt}
\end{table}



\noindent \textbf{Non-Linear Attention}. Figure \ref{fig:zipformerblock} shows the structure of NLA. Firstly, it maps the input features to A, B, and C using linear layers. Then, the NLA organizes the output as $O = linear(A \odot attention(tanh(B) \odot C))$, where $\odot$ denotes element-wise multiplication, and attention uses the attention weights provided by MHAW to organize $tanh(B) \odot C$ in the time or frequency dimension. Finally, a linear layer is used to output the final feature. 


\noindent \textbf{Bypass}. The Bypass Module, a novel form of residual connection, combines the input $x$ and output $y$ of the wrapped intermediate module using a channel-wise learnable weight $c$. The output is defined as $O = (1 - c) \odot x + c \odot y$. A smaller value of $c$ indicates a greater tendency to skip the wrapped intermediate module. During training, $c$ is initially constrained to the range of [0.9, 1.0] for the first 2k steps and then to the range of [0.2, 1.0] after 2k steps.

\noindent \textbf{BiasNorm}. Due to LN possibly introducing a large fixed value to a specific channel and causing excessive small activations in some modules, BiasNorm is used as a replacement. Its formulation is as follows: $BiasNorm(x) = \frac{x}{RMS[x - b]} \cdot exp(\gamma)$, 
where $b$ represents the learnable channel-wise bias, $RMS[x - b]$ denotes the root-mean-square value calculated over the channels, with $\gamma$ as a scalar. Using BiasNorm with ScaleAdam can achieve more stable model convergence.

\begin{table}[t]
\small

\setlength{\tabcolsep}{0.5mm}{
  \centering
  \footnotesize
  \caption{\label{tab:dns} {Comparison with other methods on the DNS Challenge 2020 test set without reverberation. $*$ represents our reproduced model. Para. and NB represent Parameters and NB-PESQ.
}}
  \begin{tabular}{ccccccccccccc}
  \toprule

  Model & Para.[M] & FLOPS[G] & WB-PESQ & NB & STOI & SI-SDR  \\

  \midrule

  FRCRN\cite{FRCRN} & 6.9 & - & 3.23 & 3.60 & 97.69 & 19.78 \\

  MFNet\cite{MFNet} & - & - & 3.43 & 3.74 & 97.98 & 20.31 \\

  TridentSE-L\cite{TridentSE} & 3.03 & 59.8 & 3.44 & - & 97.86 & - \\

  USES\cite{USES} & - & - & 3.46 & - & 98.1 & \textbf{21.2} \\

  MP-SENet Up.\cite{mpsenetv2} & 2.26 & 81.33$^{*}$ & 3.62 & 3.92 & 98.16 & 21.03 \\

  \midrule

  TF-Locoformer\cite{tflocoformer} & 14.97 & 497.24 & 3.72	& - &	\textbf{98.8} & \textbf{23.3} \\
  
  \midrule

 MP-SENet Up.$^{*}$ & 2.26 & 81.33 & 3.61 & 3.95 & 98.08 & 20.55 \\

  ZipEnhancer(S) & 2.04 & 62.85 & \textbf{3.69} & \textbf{3.99} & \textbf{98.32} & \textbf{21.15} \\

  ZipEnhancer(M) & 11.34 & 266.96 & \textbf{3.81} & \textbf{4.08} & 98.65 & 22.22 \\

  \bottomrule

\end{tabular}
}
\vspace{-15pt}
\end{table}

\begin{table*}[t]\centering
\footnotesize
    \caption{\label{tab:voice} {Comparison with other methods on VoiceBank+DEMAND dataset. $*$ represents the results of our reproduced model. Para., w/o, Up. and DP represent Parameters, without, Updated and Dual-Path.
  }}
\scalebox{1.0}{
    \begin{tabular}{cccccccccccccccccccccccccccc}
    \toprule

    Model & Year & DP & Para.[M] &	FLOPS[G] & WB-PESQ & CSIG & CBAK & COVL & STOI & SSNR & SI-SDR \\
    
    \midrule

    DB-AIAT\cite{DBAIAT} & 2022 & - & 2.81 & 68.0 & 3.31 & 4.61 & 3.75 & 3.96 & 95.6 & \textbf{10.79} 	\\

    CMGAN\cite{CMGAN} & 2022 & $\checkmark$ & 1.83 & 116 & 3.41 & 4.63 & 3.94 & 4.12 & 96 & - & -	\\

    TridentSE-L\cite{TridentSE} & 2023 & $\checkmark$ & 3.03 & 59.8 & 3.47 & 4.70 & 3.81 & 4.10	& 96 & - & - \\

    HATFANet\cite{HATFANet} & 2024 & $\checkmark$ & 3.9 & - & 3.37 & 4.66 & 3.85 & 4.11 & 95.8 & 10.15	\\

    SE-LMA-Transformer\cite{selmatransformer} & 2024 & $\checkmark$ & 5.60	& - & 3.40 & 4.65 & 3.87 & 4.12 & 95.7 & 10.15 & - \\

    Spiking-S4\cite{spikings4} & 2024 & - & 0.53 & 1.50 & 3.39 & 4.92 & 2.64 & 4.31 & - & - & -	\\

    MP-SENet\cite{mpsenet} & 2023 & $\checkmark$ & 2.05 & 77.79 & 3.45 & 4.73 & 3.95 & 4.22 & 96 & 10.64 & 19.94	\\

    UPB-CMGAN\cite{UPBCMGAN} & 2024 & $\checkmark$ & 1.83 & 116 & 3.55 & 4.78 & - & 4.28 &  96 & - & -	\\

    SEMamba w/o PCS\cite{semamba} & 2024 & $\checkmark$ & 	2.25 & 65.46 & 3.55 & 4.77 & 3.95 & 4.26 & 96 & - & -	\\

    MP-SENet Up.\cite{mpsenetv2} & 2024 & $\checkmark$ & 2.26 & 81.33$^{*}$ & 3.60 & \textbf{4.81} & \textbf{3.99} & 4.34 & 96 & - & 19.47	\\

    \midrule
    
    ZipEnhancer(S,$\lambda_{6}=0$) & 2024 & $\checkmark$ & 2.04 & 62.85 & \textbf{3.63} & \textbf{4.81} & 3.87 & \textbf{4.36} & 96.19 & 8.33 & 19.09 \\

    ZipEnhancer(S,$\lambda_{6}=0.2$)  & 2024 & $\checkmark$ & 2.04 & 62.85 & 3.61 & \textbf{4.81} & 3.97 & 4.35 & \textbf{96.22} & 10.01 & \textbf{19.96} \\


    \bottomrule
\end{tabular}
}
\vspace{-12pt}
\end{table*}

\subsubsection{Encoder and Decoders}

We use an Encoder and Decoders similar to MP-SENet\cite{mpsenet}. The Encoder encodes $Y_{in}$ to $Y^{'} \in R^{C \times T \times F^{'}}$, where $F^{'}=\frac{F}{2} + 1$. It consists of two convolutional layers with Conv2d, InstanceNorm (IN), and PRelu, along with a Dilated DenseNet. The second convolutional layer features a stride of 2 to reduce $F$ to $F^{'}$. The dilated DenseNet includes four convolutional layers with dilation sizes of 1, 2, 4, and 8, expanding the receptive field along the time axis.



The Decoders consist of a Magnitude Decoder and a Phase Decoder. The Magnitude Decoder includes a Dilated DenseNet, convolutional layers with Sub-pixel Conv2d\cite{subpixel}, IN, PRelu, and an output layer Conv2d. In this setup, LSigmoid is removed, and Conv2d is used directly to map the predicted compressed clean magnitude spectrum $\hat X_{m}^{c}$. The Phase Decoder includes a Dilated DenseNet, convolutional layers with Sub-pixel Conv2d, IN, PRelu, and two Conv2d layers with an Arctan2 component. The Phase Decoder predicts the enhanced wrapped phase spectrum $\hat X_{p}$.

\vspace{-3pt}
\subsection{Training criteria}

\subsubsection{Optimizer and Learning Sceduler}

We use ScaleAdam\cite{zipformer}, a parameter-scale-invariant version of Adam, to achieve faster convergence and improved performance. ScaleAdam scales the parameter updates based on the parameter scale, ensuring relatively consistent parameter changes for different scales, in contrast to Adam where the gradient scale for each parameter remains constant. It learns the scale of parameters by adding a gradient that scales the learning concerning the parameter scale to the original gradient, allowing each module to more easily adjust activation values to an appropriate range.
We also use Eden\cite{zipformer} as our learning rate (LR) scheduler, controlled by parameters such as $\alpha_{base}$, $t_{warmup}$, $\alpha_{step}$, and $\alpha_{epoch}$. Here, $\alpha_{base}$ is the maximum LR without warmup, $t_{warmup}$ is the number of steps for linear warmup, and $\alpha_{step}$ and $\alpha_{epoch}$ determine when the LR starts to decrease rapidly.

\subsubsection{Loss function}

The loss function used for training ZipEnhancer follows \cite{mpsenetv2}. It is a linear combination of various losses, including PESQ-based GAN discriminator $\mathcal{L}_{pesq}$, STFT consistency $\mathcal{L}_{stft}$, magnitude $\mathcal{L}_{mag}$, complex $\mathcal{L}_{com}$, and phase $\mathcal{L}_{pha}$ losses. We reuse the time loss $\mathcal{L}_{time}$ from \cite{mpsenet}. The final loss is defined as: $\mathcal{L} = \lambda_{1} \mathcal{L}_{pesq} + \lambda_{2} \mathcal{L}_{stft} + \lambda_{3} \mathcal{L}_{mag} + \lambda_{4} \mathcal{L}_{com} + \lambda_{5} \mathcal{L}_{pha} + \lambda_{6} \mathcal{L}_{time}$.




\section{Experiments}

\subsection{Dataset}

In this experiment, we validate the performance of our SE model using two widely used open-source datasets: the DNS Challenge 2020 (DNS2020) dataset and the VoiceBank+DEMAND dataset. The DNS2020 dataset\cite{dns2020} includes 500 hours of clean audio clips from 2150 speakers and over 180 hours of noise audio clips. The clean audio clips are split into 10-second segments, and 3,000 hours of noisy audio clips with SNRs ranging from -5dB to 15dB are generated for training, following the official script\footnote{https://github.com/microsoft/DNS-Challenge}. For model evaluation, the dataset provides non-blind test sets containing 150 noisy-clean pairs generated from audio clips spoken by 20 speakers. 

The VoiceBank+DEMAND dataset\cite{voicebankdemands} contains pairs of clean and noisy audio clips with a sampling rate of 48 kHz. The clean audio set is selected from the Voice Bank corpus\cite{voicebank}, which includes 11,572 audio clips from 28 speakers for training and 872 audio clips from 2 unseen speakers for testing. These clean audio clips are mixed with 10 types of noise (8 from the DEMAND database\cite{demands} and 2 artificial types) at SNRs of 0-15 dB for the training set, and 5 types of unseen noise from the DEMAND database at SNRs of 2.5, 7.5, 12.5, and 17.5 dB for the test set. In the experiments, we resample all the audio clips to 16 kHz.

\vspace{-3pt}
\subsection{Expermental Setup}


During training, all audio clips are segmented into 2-second segments. To extract input features using STFT from the raw waveform, the parameters for FFT point number, Hanning window size, and hop size are set to 400, 400 (25 ms), and 100 (6.25 ms), resulting in the number of frequency bins, $F = 201$. The magnitude spectrum compression factor $c$ is set to 0.3. For training ZipEnhancer, we define the batch size (B) to 4. Details about the settings for the number of FT-Zipformerblocks (N), number of channels (C), the number of heads in the MHAW, and the down-sampling ratios can be seen in Table \ref{tab:config}.
We can conveniently design models at different computational levels by using different down-sampling ratios, allowing for adaptation to various inference conditions and achieving a trade-off between performance and computational costs. The hyper-parameters of the final generator loss $\lambda_{1}, \lambda_{2}, ..., \lambda_{6}$ are set to 0.05, 0.1, 0.9, 0.1, 0.3, and 0.2. All models are trained for 600k steps using the ScaleAdam optimizer. The hyper-parameters of the Eden scheduler, $\alpha_{base}$, $t_{warmup}$, $\alpha_{step}$, and $\alpha_{epoch}$, are set to 0.04, 4000, 2500, and 24. Training is performed on one 32GB V100 RTX GPU. For training ZipEnhancer(M), the audio clips are segmented into 4-second segments to ensure a fair comparison with TF-Locoformer\cite{tflocoformer}, the model is trained for 500k steps using four 32GB V100 RTX GPUs.

\vspace{-3pt}
\subsection{Evaluation metrics}

For the evaluation on the DNS2020 dataset, four commonly used objective evaluation metrics are chosen to evaluate the enhanced speech quality, including wide-band PESQ (WB-PESQ), narrowband PESQ (NB-PESQ), short-time objective intelligibility (STOI) and the scale-invariant signal-to-distortion ratio (SI-SDR)\cite{sisdr} to quantify the distortion between the enhanced and clean speech signals. For VoiceBank+DEMAND dataset, besides the WB-PESQ, STOI and SI-SDR, we use the segmental SNR (SSNR) and three composite measures (CSIG, CBAK, and COVL) which predict the mean opinion score (MOS) on signal distortion, background noise intrusiveness, and overall effect, respectively.
For all the metrics, higher values indicate better performance. 


\subsection{Expermental Results}

\begin{figure}[!t]
	\centering
	\includegraphics[scale=1.1]{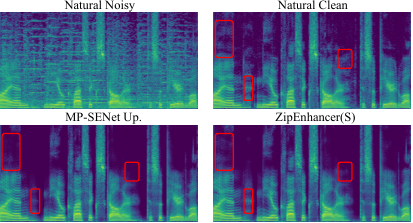}
        \vspace{-7pt}
	\caption{
		Spectrogram visualization of the natural noisy/clean speech, and speeches enhanced by the MP-SENet Up. and our proposed ZipEnhancer(S).
	}
	\label{fig:spec}
        \vspace{-10pt}
\end{figure}

\begin{table}[t]\centering
\small

\setlength{\tabcolsep}{0.8mm}{
  \centering
  \footnotesize
  \caption{\label{tab:ablation} {Results of different model configurations and Ablation study on the DNS Challenge 2020 test set without reverberation.
}}
  \begin{tabular}{ccccccccccccc}
  \toprule

  Model & FLOPS[G] & WB-PESQ & NB-PESQ & STOI & SI-SDR & SSNR \\

  \midrule

  S & 62.85 & \textbf{3.69} & \textbf{3.99} & 98.32 & 21.15 & 15.06 \\
  \midrule
  S2 & 80.61 & 3.67 & 3.98 & \textbf{98.34} &	\textbf{21.35} & \textbf{15.14} \\
  \midrule
  S3 & 60.79 & 3.68 & \textbf{3.99} & 98.31 & 21.17 & 15.00 \\
  S4 & 51.91 & 3.65 & 3.98 & 98.20 & 20.76 & 14.81 \\
  S5 & 49.84 & 3.64 & 3.96 & 98.18 & 20.71 & 14.49 \\
  S6 & 41.47 & 3.60 & 3.95 & 98.11 & 20.53 & 14.62 \\
  S7 & 40.06 & 3.58 & 3.93 & 98.06 & 20.36 & 13.45 \\
  S8 & 36.94 & 3.54 & 3.91 & 97.90 & 19.83 & 13.53 \\

  \midrule

  S(AdamW) & 62.85 & 3.65 & 3.97 & 98.19 & 20.84 & 14.75 \\
  




  \bottomrule

\end{tabular}
}
\vspace{-15pt}
\end{table}

\subsubsection{Comparison with other state-of-the-art models}

Evaluation on the DNS2020 test set without reverberation: In Table \ref{tab:dns}, TridentSE-L\cite{TridentSE}, USES\cite{USES}, MP-SENet Updated\cite{mpsenetv2} and TF-Locoformer\cite{tflocoformer} are all Dual-Path models. Our ZipEnhancer(S) outperforms these recent SOTA models on several metrics, except TF-Locoformer. To further analyze the impact of model complexity, we reproduce the results of MP-SENet Up.($\lambda_{6}=0.2$)$^{*}$, achieving comparable results to \cite{mpsenetv2} with approximately 81.33 FLOPS. Our model reaches a new SOTA PESQ of 3.69 while maintaining a lower parameter count than recent models and much lower or comparable FLOPS. This indicates the superiority of our current model and the effectiveness of the Down-Up Sampling structure in reducing computational costs. We also visualized the spectrograms of speeches as shown in Fig. \ref{fig:spec}. Additionally, our ZipEnhance(M) achieves performance similar to that of TF-Locoformer with only half the FLOPS.

Evaluation on VoiceBank+DEMAND dataset: In Table \ref{tab:voice}, our ZipEnhance(S, $\lambda_{6}=0$) achieves a new SOTA result in WB-PESQ on the current dataset, with a parameter count of 2.04 and lower FLOPS of 62.85. Compared to MP-SENet Up., our model improves several metrics, including WB-PESQ, CSIG, COVL, and STOI. Our model achieves promising results with a lower parameter count and reduced computational costs, demonstrating the superiority of the ZipEnhancer. Additionally, according to \cite{cmganv2}, $\mathcal{L}_{time}$ can improve SI-SDR but may reduce WB-PESQ. We also implement ZipEnhance(S, $\lambda_{6}=0.2$), which achieves better PESQ than  MP-SENet Updated while improving SI-SDR metrics. Audio samples processed by the proposed ZipEnhancer on the DNS2020 and VoiceBank+DEMAND datasets can be accessed at the website.\footnote{https://zipenhancer.github.io/ZipEnhancer/} 

\subsubsection{Results of different model configurations}

In Table \ref{tab:ablation}, we compare the results of different model configurations on the DNS2020 test set without reverberation. When setting all down-sampling ratios to 1, we obtain S2. S maintains high performance and gets metrics similar to those of S2, showing that the efficient down-sampling module does not cause much information loss. S3-S8 use more aggressive down-sampling ratios, resulting in lower FLOPS and gradually decreasing performance while still maintaining PESQ above a specific threshold (3.54, better than \cite{USES,TridentSE}), demonstrating the ability to achieve a trade-off between computational cost and performance. Additionally, we conduct an ablation experiment on ScaleAdam, replacing it with the original LR of 0.0005 for AdamW in conjunction with the LR scheduler of \cite{mpsenetv2}. The performance shows a slight decrease, indicating the effectiveness of ScaleAdam.

\vspace{-1pt}
\section{Conclusion}
\vspace{-1pt}
Compared to other sequence tasks that only model the hidden layer features with three axes, Dual-Path time and time-frequency domain speech enhancement models, while effective and with a small parameter count, entail high computational costs due to their  hidden layer features with four axes. To address this, we propose ZipEnhancer, which incorporates time and frequency domain DownSampling to reduce computational overhead. Our model achieves new SOTA results on the DNS 2020 Challenge and Voicebank+DEMAND datasets, with a PESQ of 3.69 and 3.63, respectively, utilizing 2.04M parameters and 62.85G FLOPS. The introduction of the ScaleAdam optimizer and the Eden learning rate scheduler further contribute to the model's effectiveness. The results show the effectiveness of our ZipEnhancer, establishing our model as a competitive solution that surpasses other models with similar complexity levels. In the future, we plan to design a causal model using ZipEnhancer to facilitate real-time deployment.


\begin{spacing}{0.95}
\footnotesize
\normalem
\bibliographystyle{IEEEbib}
\bibliography{refs}
\end{spacing}

\end{document}